# Candidate Auroral Observations during the Major Solar-Terrestrial Storm in May 1680: Implication for Space Weather Events during the Maunder Minimum

Hisashi Hayakawa (1 – 3)*, Kristian Schlegel (4), Bruno P. Besser (5), Yusuke Ebihara (6 – 7)

(1) Institute for Space-Earth Environmental Research, Nagoya University, Nagoya, 4648601, Japan

(2) Institute for Advanced Researches, Nagoya University, Nagoya, 4648601, Japan

(3) RAL Space, Rutherford Appleton Laboratory, Science and Technology Facilities Council, Harwell Campus, Didcot, OX11 0QX, UK

(4) Copernicus Gesellschaft e.V, Bahnhofsallee 1e, 34081 Göttingen, Germany.

(5) Space Research Institute, Austrian Academy of Sciences, 8042 Graz, Austria

(6) Research Institute for Sustainable Humanosphere, Kyoto University, Uji, 6110011, Japan

(7) Unit of Synergetic Studies for Space, Kyoto University, Kyoto, 6068306, Japan

* hisashi@nagoya-u.jp

**Abstract**

The Maunder Minimum (1645–1715) is currently considered the only grand minimum within telescopic sunspot observations since 1610. During this epoch, the Sun was extremely quiet and unusually free from sunspots. However, despite reduced frequency, candidate aurorae were reported in the mid-European sector during this period and have been associated with occurrences of interplanetary coronal mass ejections (ICMEs), whereas some of them have been identified as misinterpretations. Here, we have analysed reports of candidate aurorae on 1 June 1680 with simultaneous observations in mid-Europe, and compared their descriptions with visual accounts of early modern aurorae. Most contemporary sunspot drawings from 22, 24, and 27 May 1680 have shown that this apparent sunspot may have been a source of ICMEs, which caused the reported candidate aurorae. On the other hand, its intensity estimate shows that the magnetic storm during this candidate aurora was probably within the capability of the storms derived from the corotating interaction region (CIR). Therefore, we accommodate both ICMEs and CIRs as their possible origin. This interpretation is probably applicable to the candidate aurorae in the often-cited Hungarian catalogue, on the basis of the reconstructed margin of their equatorward auroral boundary. Moreover, this catalogue itself has clarified that the considerable candidates during the MM were probably misinterpretations. Therefore, frequency of the auroral visibility in Hungary was probably lower than previously considered and agree more with the generally slow solar wind in the existing reconstructions, whereas sporadic occurrences of sunspots and coronal holes still caused occasional geomagnetic storms.

**1. Introduction**

Among the coverage of direct solar observations for the last four centuries, the Maunder Minimum (MM; 1645–1715) was the only grand minimum characterised with extremely suppressed solar cycles, asymmetric sunspot occurrences, and probable loss of significant streamers from the solar corona (Eddy, 1976; Ribes and Nesme-Ribes, 1993; Riley *et al*., 2015; Vaquero *et al*., 2015; Usoskin *et al*., 2015; Owens *et al*., 2017). Such characteristics are quite unique within the coverage of direct solar observations, including the Dalton Minimum and the usual cycle minima (Clette *et al*., 2014; Hathaway, 2015; Owens *et al*., 2017; Muñoz-Jaramillo and Vaquero, 2019; Hayakawa *et al*., 2020a, 2020b) and associated with a special state of the solar-dynamo behaviour (*e.g.* Charbonneau, 2020). Being the only grand minimum within the coverage of direct solar observations, the MM forms a reference for other grand minima confirmed in the proxy reconstructions based on cosmogenic isotopes (Usoskin *et al*., 2007; Muscheler *et al*., 2007, 2016; Inceoglu *et al*., 2015; Usoskin, 2017; Wu *et al*., 2018). Accordingly, open solar flux and the interplanetary magnetic field were probably weakened more than during the normal solar minima, although their exact amplitude are still under discussion (Beer *et al*., 1998; Cliver and Ling, 2011; Owens and Lockwood, 2012; Lockwood, 2013; Lockwood and Owens, 2014; Cliver *et al*., 2013; Svalgaard, 2013; Usoskin *et al*., 2015, 2017; Vaquero *et al*., 2015; Zolotova and Ponyavin, 2015, 2016; Svalgaard and Schatten, 2016; Owens *et al*., 2017).

This variability probably influenced the frequency of solar eruptions as well, although the significance of this influence has not yet been clearly determined. Statistical analyses imply that the occurrence rate of interplanetary coronal mass ejections (ICMEs) was somewhat independent and probably comparable between recent solar minima (2008/2009 and 1996/1997) and the MM (Owens and Lockwood, 2012). In fact, the occurrences of large ICMEs and magnetic storms have been shown without exact correlation with sunspot number (Kilpua *et al*., 2015; Lefèvre *et al*., 2016); some extreme ICMEs and geomagnetic storms are known to occur even around the deep solar minima (Garcia and Dryer, 1987; Daglis *et al*., 2007; Hayakawa *et al*., 2020c).

It is challenging to directly track solar eruptions during the MM, given its occurrence far before the onset of geomagnetic observations (Usoskin *et al.*, 2015). Nevertheless, major solar eruptions leave a footprint as mid- to low-latitude aurorae, if they cause a long-lasting southward interplanetary magnetic field (IMF) in ICMEs, and/or their front-side sheath region, resulting in geomagnetic storms (Gonzalez *et al.*, 1994; Daglis *et al.*, 1999; Tsurutani *et al.*, 2003; Cliver and Dietrich, 2013). Such aurorae have been recorded for millennia in historical documents (Siscoe, 1980; Silverman, 1992, 1998; Stephenson et al., 2004; Vaquero and Vazquez, 2009; Schlegel and Schlegel, 2011; Hayakawa *et al.*, 2017, 2019b). Archival investigations show that the candidate aurorae seemed to be reported in the European sector during the MM (Eddy, 1976, 1983; Mendillo and Keady, 1976; Link, 1977; Schröder, 1978, 1988, 1992; Siscoe, 1980; Feynman and Gabriel, 1990; Schlamminger, 1990, 1991; Legrand *et al.*, 1991; Silverman, 1992, 1993, 1998; Letfus, 2000; Lockwood and Barnard, 2015; Riley *et al.*, 2015; Zolotova and Ponyavin, 2015, 2016; Usoskin *et al.*, 2015, 2017; Vázquez *et al.*, 2016; Ogurtsov, 2019). Some of these studies have arguably highlighted their reduced frequency and few pairs with the observation of source sunspots but associated them with ICMEs based on their latitudinal distributions (Letfus, 2000; Riley *et al.*, 2015; Usoskin *et al.*, 2015; Vázquez *et al.*, 2016).

However, a caveat must be noted that the interactions of the high-speed solar-wind streams with the upstream slow-speed streams can generate corotating interaction regions (Smith and Wolfe, 1976; Tsurutani *et al.*, 1995; Richardson *et al.*, 2002, 2006; Gopalswamy *et al.*, 2015) and cause moderate geomagnetic storms and aurorae (*e.g.* Usoskin *et al.*, 2015; Vázquez *et al.*, 2016). As the majority of such mid-latitude aurorae were reported without plausible source sunspots and remained in medium geographic latitudes, it is still controversial how many of these aurorae resulted from solar eruptions (Letfus, 2000; Zolotova and Ponyavin, 2016; Usoskin *et al.*, 2017).

Furthermore, careful analysis of the original historical records is needed to assess the reliability and magnitude of these reports on candidate aurorae. On the one hand, it has been confirmed that some candidate aurorae are probably misinterpretations of other phenomena such as atmospheric optics and hence are excluded from discussions on the auroral activity during the MM (*e.g.* Rethly and Berkes, 1963; Kawamura *et al.*, 2016; Usoskin *et al.*, 2017). On the other hand, the extent of the equatorial auroral boundary shows a fairly good correlation with the intensity of the associated geomagnetic storms (Yokoyama *et al.*, 1998) and hence, should be reconstructed based on the reported details (*e.g.* Hayakawa *et al.*, 2018). Their origin could be inferred on the basis of contemporary sunspot observations (*e.g.* Letfus, 2000; Willis *et al.*, 2005), and compared to reconstructed storm intensity with the observed threshold of storms derived from the corotating interaction region (*e.g.* Richardson *et al.*, 2006).

In this context, it is important to analyse likely robust geomagnetic storms in the core MM during the period 1650–1700 (Vaquero and Trigo, 2015; Vaquero *et al.*, 2015; *c.f.*, Svalgaard and Schatten, 2016). One such case was a major storm in March 1653 confirmed by simultaneous observations in East Asia (Willis and Stephenson, 2000; Isobe *et al.*, 2019). Another candidate case with simultaneous observations is known from May 1680 (Fritz, 1873; see also Schröder, 1978). Here, we examine its reported details from multiple observations in Central Europe to assess their reliability and reconstruct its spatial extent. We also investigate contemporary solar observations around this event and consider their plausible solar source. On their basis, we empirically infer both storm magnitude and source, to derive further implications for space weather variability during the MM.

**2. Interpretation of the fire-sign on 1 June 1680**

The original source documents for the simultaneous candidate aurorae are found in a journal *Neue Himmels Zeitung* (Kirch, 1681), which Gottfried Kirch (1639 – 1710) compiled. Close inspection reveals that Kirch (1681) probably owes his reports around Hamburg to *Planeten-Versamblung im Majo und Junio 1680* (Voigt, 1681). The recorded signs in Kirch were summarised as "the great fire-sign, which appeared at many places in Germany, particularly at Leipzig/Hamburg/Lübeck and other places on 22 May in the Sky early in the day" and interpreted as candidate aurorae in Fritz (1873) and Schröder (1978). In his publication, Kirch has collected the reports for this "fire-sign" from witnesses, whereas he himself missed the display. Reports around Hamburg were collected in *Planeten-Versamblung im Majo und Junio 1680*, a German pamphlet (Voigt, 1681), which the astronomer Johann Heinrich Voigt (1613 – 1691) compiled. These records are cited into Kirch (1681). As the Julian calendar was in use in Germany before 1700, this date should be converted to 1 June in the Gregorian calendar (*e.g.* Von Aufgebauer, 1969).

According to Kirch (1681) and Voigt (1681), this fire-sign was reported widely around the Western Coast of the Baltic Sea and Leipzig (Figure 1) "early in the day" on 1 June, namely during the night between 31 May and 1 June 1680. The longest duration of the fire-sign was reported at Leipzig, between 1 LT and daybreak (3:48 LT), from multiple witnesses. With variable onsets, this fire-sign persisted up to 3 LT–4 LT (Kirch, 1681, pp. 3–5; Voigt, 1681, p. 1), which is consistent with the computed timing of local daybreaks (*e.g.* 3:34 LT at Lübeck, and 3:25 LT at Nyborg).

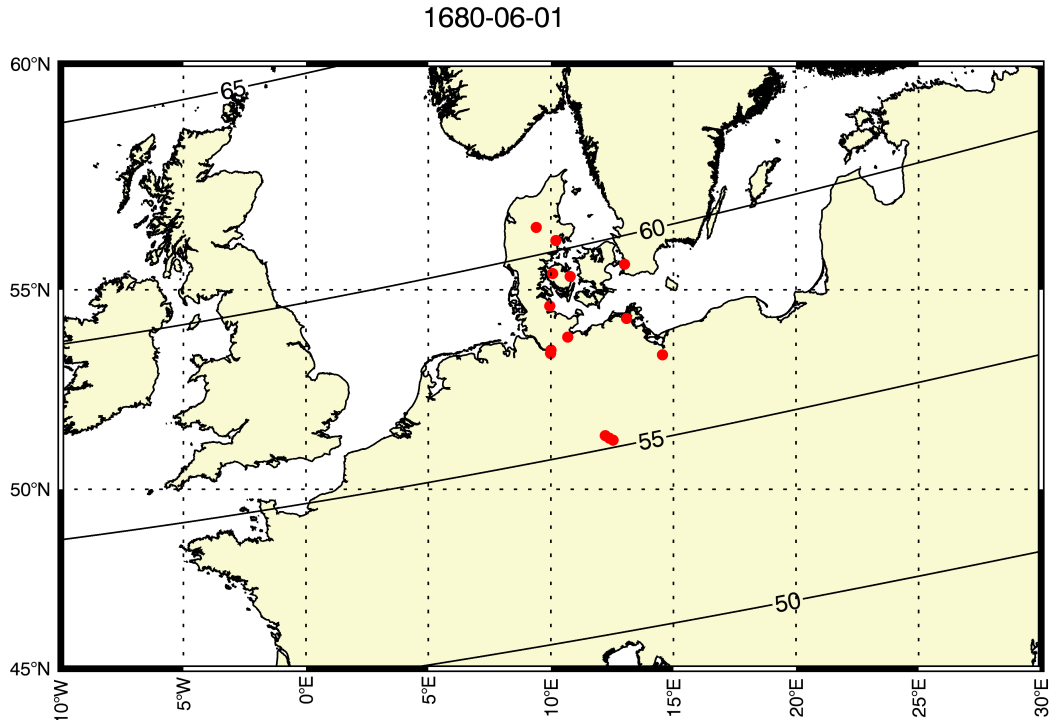

Figure 1: Observational sites of the 'fire-sign' in comparison with the contours of the magnetic latitude (MLAT) for every 5°, based on the archaeo-magnetic field model Cals3k4b (Korte and Constable, 2011).

This sign seemed to occupy a large part of the sky, especially northward. Haarburg (N53°28′, E09°59′) witnessed rays "rising like a lightning flash against NE, or in fact NNE" (Kirch, 1681, p. 7). At Leipzig (N51°20′, E12°23′) its extension was witnessed probably overhead, reporting included the statement "the whole sky which I could see towards west was filled with fiery mist" and "towards East, it was beautiful blue and brightly starred" (Kirch, 1680, p. 3). Over Carlsburg (N54°37′, E09°57′), a witness stated "in all four directions of the sky appeared a great number of folks with different figures and clothing among which those of the NNE persisted the longest time" (Kirch, 1680, pp. 6–7).

This fire-sign was described as glow and strokes with various colours (reddish, bluish, golden, and silvery). At Leipzig, it was described as "reddish or fiery fog" and "fiery mist" (Kirch, 1681, p. 3). At Haarburg, the reported rays were "a snake head and the rays shooting against, like vʌvʌv could well have presented a figure like a curved snake" (Voigt, 1681, p. 3). At Hamburg, "On top there was a bright star, wherefrom the fire or the rays were broader and downwards they become smaller, like angles and tips, like ᴠʌᴠʌᴠ against each other staying" (Voigt, 1681, p. 3). Overall, its shape was variably described as rays, fiery cloud, fire fall, open of sky, serpent, long beams, cloths with variable letters, and balls with curved trails (Voigt, 1681, pp. 2–4; Kirch, 1681, pp. 3–5). Its shape varied in time, as shown in Figure 2 (from N° 1 to N° 3) for a specific case at Lübeck (N53°52′, E10°41′), for example. Contemporaries at Lübeck heard "some noise and strong bangs from shooting" (Voigt, 1681, p. 5), "as if a rocket would be in the air" and "as if a musket was fired" (Kirch, 1681, pp. 5–6).

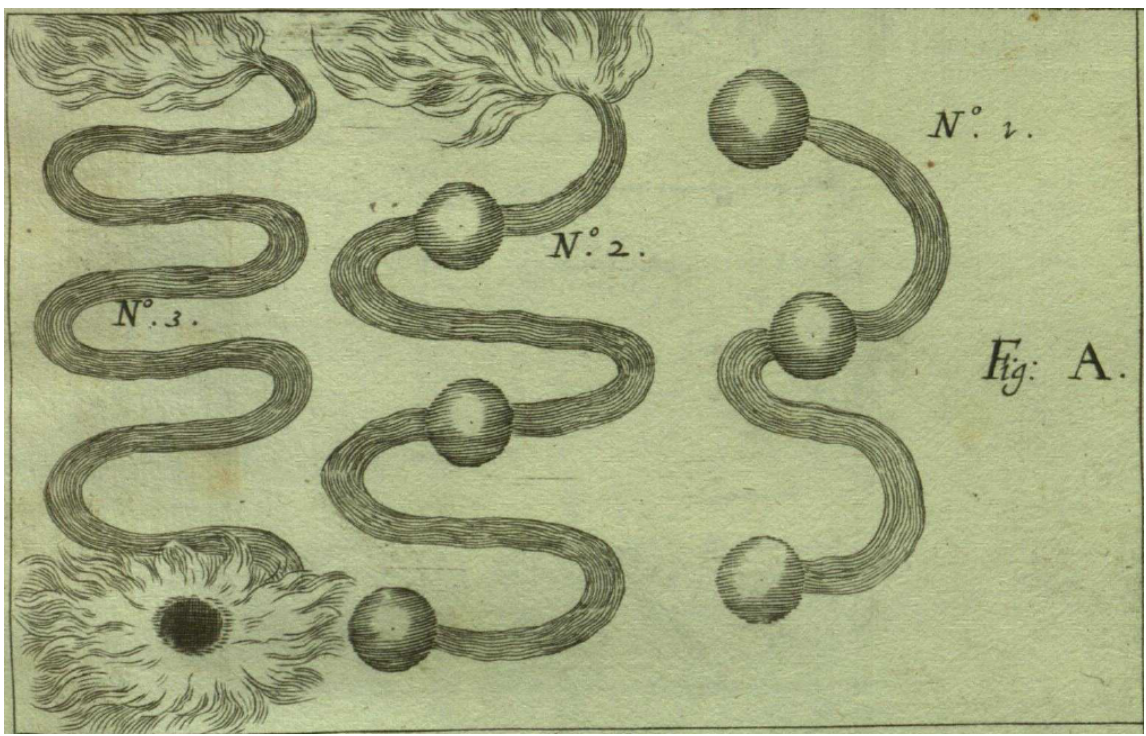

Figure 2: Depicted shape of the 'fire-sign' reported from Lübeck on 1 June 1680, reproduced from Kirch (1681).

Its interpretation is challenging, as all the reported details do not fit well together. The visual accounts favour the auroral interpretation in Fritz (1873) and Schröder (1978), whereas the acoustical details sound like a fireball. Kirch immediately rejected the contemporary proposals of the anomalous planetary motion and possible solar halo, due to its motion and observational time. Atmospheric optics would contradict the reported colourations and extents, as its colouration in the night sky is virtually invisible and they stay around the Moon (Minnaert, 1993, pp. 219–213). Fireballs seem to be the most favourable explanation, with reported sizzling like a rocket, and the bang and crackle noise. However, this description is inconsistent with the reported curved shape and long persistence, since fireballs fly linearly and do not last long. In addition, a witness at Fuchshan (probably "Fuchshain" close to Leipzig) reported a "whimsical cloud" remaining after the disappearance of bright lines. The shape depicted in Figure 2 is far from what should be expected for the trail of fireballs.

This leaves Fritz and Schröder's auroral interpretation still plausible, with the support of other simultaneous observations (see Willis and Stephenson, 2000). The strongest argument against it is the reported noise, which was heard locally from Lübeck and could be associated with often-reported auroral audibility (Silverman and Tuan, 1973). In fact, the reported directions, motions, and colourations are consistent with the behaviour of mid-latitude aurorae. The observational directions concentrated more towards the north, while the sign itself had a wider extent. The reported rays centred NE to NNE at Haarburg. The NNE part lasted the longest at Carlsburg, even though this sign appeared in all four directions of the sky. This is also the case of its reported motions, with notable variability within hours (Figure 2). The reported colourations are dominantly reddish with some parts being whitish (silvery), bluish and yellowish (golden). The reddish glow agrees with the $O_I$ emissions in 630.0 nm or SAR arcs typical with mid-latitude aurorae (Tinsley *et al*., 1984; Kozyra *et al*., 1997). The whitish glows are typically greenish emissions (557.7 nm) without enough brightness and typically seen in the ray structure (Ebihara *et al*., 2017; Stephenson *et al*., 2019; Bhaskar *et al*., 2020). The yellowish (golden) colour could be their mixture. The bluish emissions could be sunlit aurorae derived from $N_2^+$ emissions of 427.8 nm in the upper atmosphere at 600 – 1100 km (Hunten, 2003).

The event occurrence from 1–4 LT on 1 June indicates its visibility during twilight, and indicates that part of the electrons possibly precipitated into the sunlit area in the upper atmosphere and likely caused sunlit aurorae in bluish colouration (see *e.g*. Hunten, 2003). In fact, the eastward visibility of the bluish colouration at Leipzig at 1–2 LT is also consistent with this scenario, as the Sun was situated more eastward below the horizon during the morning sector. The visibility in twilight also indicates its significant brightness, as case reports confirm during major solar storms in August 1859, October 1870, February 1872, and May 1921 (Silverman and Cliver, 2001; Vaquero *et al*., 2008; Hayakawa *et al*., 2019a). Bright aurorae lasting from dawn to daybreak are also reported in more moderate storms. On 6 August 1860, aurorae were reported in New York, "From 2 A.M. to daybreak, auroral beams were observed, many of them coloured and shooting up to the zenith with occasional waves of light" (Hough, 1872, p. 311). This chronologically agrees with an occurrence of magnetic disturbance at Helsinki (Nevanlinna, 2004, 2006). The daybreak had unfortunately obscured its actual end, contrary to its onset at 1 LT.

Figure 1 shows the observational sites of this candidate aurora in comparison with the MLAT computed with the archaeo-magnetic field model Cals3k4b (Korte and Constable, 2011). This shows the equatorial extent of the auroral visibility down to 55.3° MLAT. The overhead visibility at Leipzig (55.3° MLAT) locates the footprint of the magnetic field line for the equatorial boundary of the auroral oval below 56.5° MLAT, under assumption of the auroral upper height as ~400 km (Roach *et al.*, 1960). With the aid of the empirical model suggested by Yokoyama et al. (1998), the minimal Dst is estimated to be −65 nT from this equatorward extent. However, we have to note that aurorae were seen overhead at London at 54.0° MLAT (Hallissey, 1974; Knipp *et al.*, 2018) during the major storm of 4 August 1972 (minimal Dst = −125 nT). The stronger dipole moment of the Earth in 1680 (≈ 1.2 times of the modern one; see Figure 4 of Korte and Constable (2011)) probably makes this estimate rather conservative, as this increase would require the equatorial boundary of the auroral oval put ≈ 1° MLAT poleward (Ebihara and Tanaka, 2020). As such, intensity of this candidate storm can be considered comparable to the 1972 storm (minimal Dst ≤ −100 nT). This consideration is fairly well consistent with Figure 3 of Yokoyama et al. (1998), showing that Dst ranges from −30 to −100 nT for the equatorward boundary of 56.5° MLAT.

### 3. Solar surface on 20–30 May 1680 and possible solar-origins

Interestingly, these reports of candidate aurorae chronologically coincide with the occurrence of sunspots in late May 1680, based on Gian Domenico Cassini and Gottfried Kirch's sunspot observations (Hoyt and Schatten, 1998a, 1998b; Neuhäuser *et al.*, 2018). Cassini reported visibility of a large sunspot from 20–30 May: "We observed on 20th May a large spot on the Sun it was already advanced on the disc of this star; it ceased to appear by passing over the upper Hemisphere of the Sun on the 30th of the same month" (Académie des sciences, 1733, pp. 317–318).

Kirch probably witnessed the same sunspot on 22–27 May (Hoyt and Schatten, 1998a, 1998b) and 28 May (Neuhäuser *et al.*, 2018) but not on 29 May (Hoyt and Schatten, 1998a, 1998b; Neuhäuser *et al.*, 2018). This sunspot was located in −0.6° ± 13.2° at its latitude, based on Kirch's published drawing on 22 May 1680 (Kirch, 1681, pp. 10–11; Neuhäuser *et al.*, 2018). Fortunately, our investigations at Paris Observatory located his original manuscripts with sunspot drawings on 12, 14, and 17 May 1680 in the Julian calendar, *i.e.* 22, 24, and 27 May 1680 as per the Gregorian calendar (Figure 3). This figure explicitly shows the visibility of the said sunspot in the western hemisphere.

Figure 3: Gottfried Kirch's sunspot drawings on 12, 14, and 17 May 1680 in the Julian calendar, adapted from MS B 3/1-6 Cote Delisle 77 at Paris Observatory (courtesy of l'Observatoire de Paris), with its brightness enhanced. The original dates are given in the Julian calendar; these are converted to 22, 24, and 27 May 1680 as per the Gregorian calendar. [NB: only available in the record version]

Its chronological correspondence with the candidate aurora on 1 June 1680 appears to be more than a coincidence. While ICMEs from the central meridian tend to be more geo-effective, they are still reportedly capable of causing major storms (−300 nT < minimal Dst ≤ −100 nT) in the western hemisphere (*e.g.* Figure 28 of Lefèvre *et al*., 2016; Figure 2 of Gopalswamy, 2018) and the distributions of geo-effective ICMEs have significant western bias (Gopalswamy *et al*., 2007). Given the variability of ICME transit time of 0.6–5 days from their launch to arrival at the Earth (*e.g.* Lefèvre *et al*., 2016; Chertok, 2020), the source solar eruption should be located somewhere within the 27–31 May 1680 period. Therefore, it is quite plausible to expect this sunspot as a source of the reported candidate aurorae in Kirch (1681).

However, with our intensity estimate of minimal Dst ≤ −100 nT, we cannot exclude the possibility that the storm was caused by a CIR resulting from the interaction between the high-speed solar wind and upstream slow-speed stream from its potential source (Smith and Wolfe, 1976; Tsurutani *et al*., 1995; Richardson *et al*., 2002, 2006). It is known that the CIR can drive a forward shock upon the Earth's magnetosphere. Its maximal intensity had actually been reported up to minimal Dst ≤ −161 nT in the interval 1972–1995 (Richardson *et al*., 2006) and roughly agrees with the theoretical limit of the minimal Dst ≤ −180 nT (O'Brien and McPherron, 2000). As such, we conclude both ICME and CIR could be a potential source for the candidate aurora on 1 June 1680.

**4. Frequency of candidate aurorae and variation of magnetic latitude in the European sector**
The reconstructed equatorial auroral boundary and the expected storm magnitude are much more decent than what would be expected for the auroral visibility in modern Europe. One such reason is the secular variations of the tilt angle of the dipole moment of the Earth, which results in the secular variations of the magnetic latitude of the European sector. Figure 4 shows these variations at Leipzig (N51°20′, E12°23′), Budapest (N47°30′, E019°03′), and Oxford (N51°45′, W001°15′), to represent observations in Germany (this study), Hungary (Rethly and Berkes, 1963; Vaquero and Trigo, 2015; Riley *et al*., 2015), and England (Usoskin *et al*., 2015). During the MM (1645 – 1715), these sites were located ≈ 55° MLAT, 51° MLAT, and 57° MLAT and ≈ 4° closer to the magnetic pole than the modern time: Leipzig = 51.2° MLAT, Budapest = 46.3° MLAT, and Oxford = 53.9° MLAT in 2014 with IGRF12 model (Thébault *et al*., 2015).

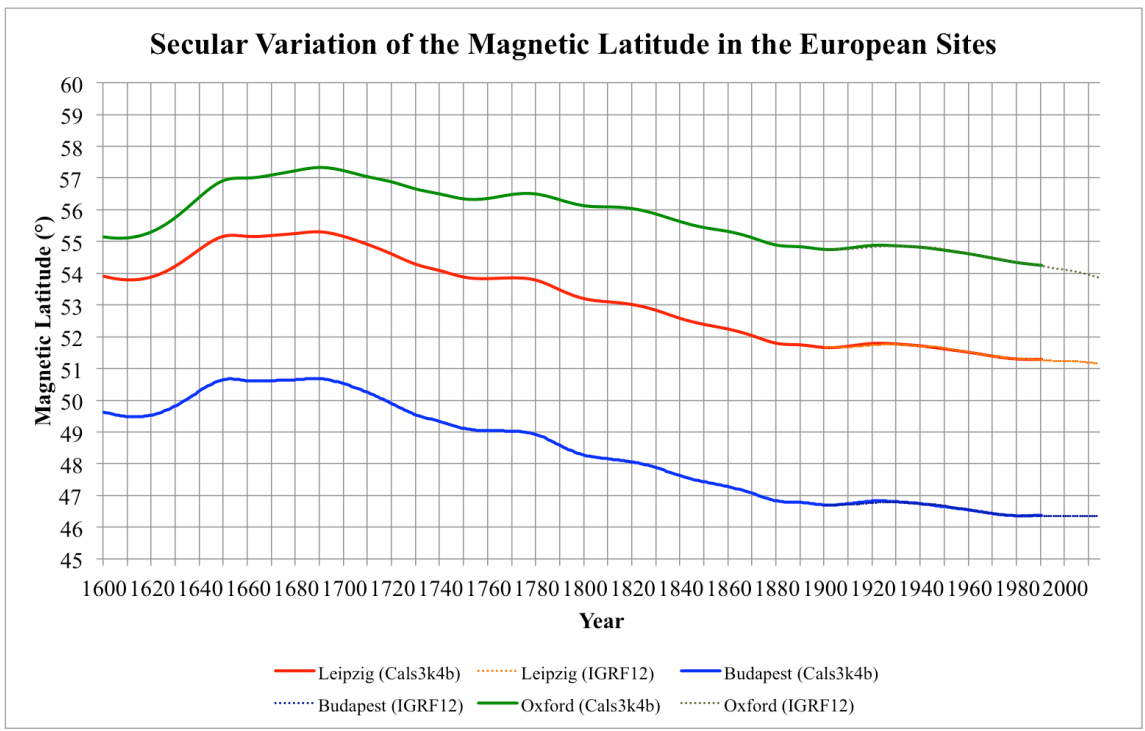

Figure 4: Secular variation of the MLATs in the European sector, represented with Leipzig in Germany (this study), Budapest Hungary (Rethly and Berkes, 1963; Vaquero and Trigo, 2015; Riley *et al*., 2015), and Oxford in England (Usoskin *et al*., 2015). The MLATs have been computed with Cals3k4b for 1600 – 1990 (Korte and Constable, 2011) and with IGRF12 for 1900 – 2014 (Thébault *et al*., 2015). These secular variations emphasise significance of apparent dearth of the candidate aurorae in England between 1621 and 1716 is especially notable, given its proximity to the magnetic pole suggested in Usoskin *et al*. (2015).

Therefore, if we assume a similar level of solar activity compared to modern times, the auroral nights would be significantly more frequent than at modern times because the European MLATs during the MM had been higher and therefore relatively closer to the auroral zone (see *e.g*, Bond and Jacka, 1962). However, the Hungarian auroral catalogue by Rethly and Berkes (1963), which is frequently cited due to its better homogeneity (Scafetta and Wilson, 2013; Vaquero and Trigo, 2015; Riley *et al*., 2015), shows a significant decrease of the nights with candidate aurorae during the MM (Figure 2 of Vaquero and Trigo, 2015; Figure 1 of Riley *et al*., 2015). Furthermore, Rethly and Berkes (1963, pp. 44–48) themselves had explicitly clarified that five of the 12 candidates (*i.e.* 1660, 1663, 1664, 1687, and 1705) were probably misinterpretations of other phenomena such as haloes or fireballs in their own notes, and one candidate was observed around sunset (1687), as summarised in Figure 5. These clarifications in Rethly and Berkes (1963) caution us towards further possible misinterpretations of the existing candidate auroral records in mid-Europe during the MM and indicate the actual frequency of the auroral night even lower than previously considered. These contrasts strongly support the peculiarity of the MM and its significant decrease in the geomagnetic activity (*e.g.*, Usoskin *et al*., 2015).

We should consider the secular variation of the strength of the dipole moment of the Earth. When the dipole moment of the Earth increases, the equatorward boundary of the auroral oval moves poleward, according to a theoretical study (Siscoe and Christopher, 1975) and a simulation study (Ebihara and Tanaka, 2020). This implies that stronger dipole moment of the Earth during the MM may have required slightly stronger magnetic storm (and the solar wind conditions) to realise the auroral oval in the same geographical extent. The simulation results obtained by Ebihara and Tanaka (2020) demonstrate that the influence of the strength of the dipole moment on the equatorward boundary of the auroral oval could be small, at least, for the past 1000 years.

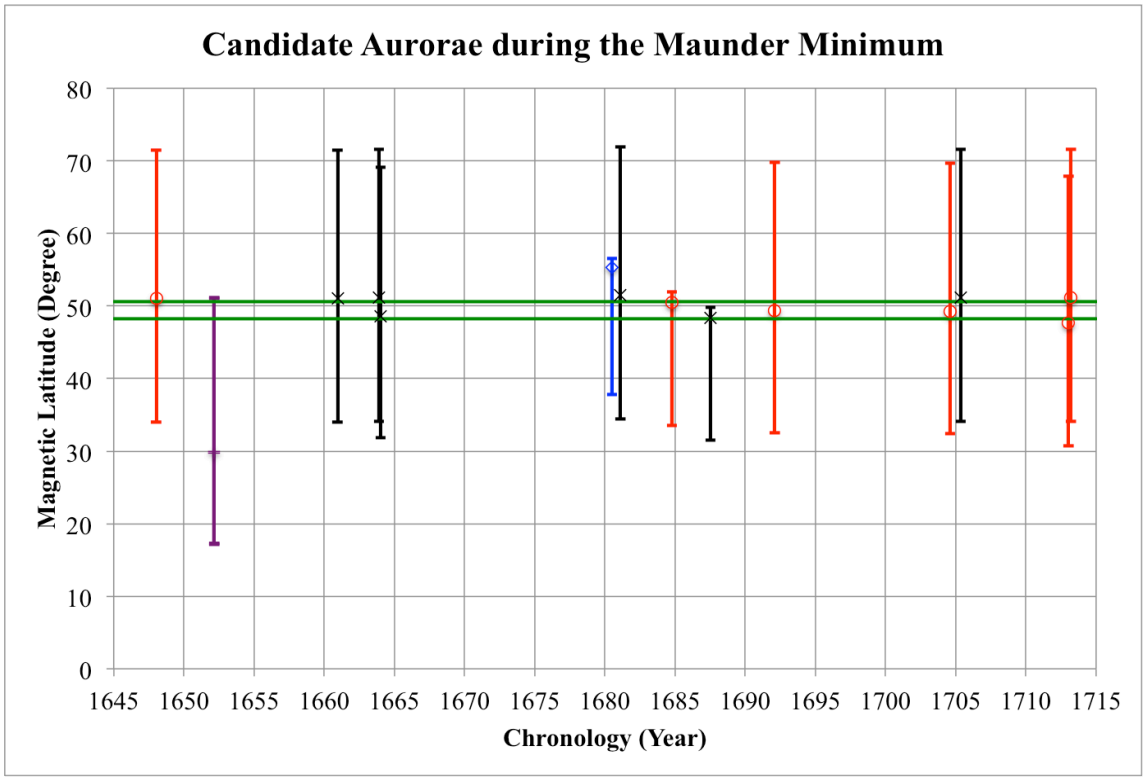

Figure 5: Estimated threshold variation (49.4° ± 1.2° MLAT) of the footprint of the magnetic field lines of the equatorial auroral boundary during the CIR-storms (green horizontal lines) based on Yokoyama et al. (1998) *vs* estimated ranges of the equatorial boundaries of the candidate aurorae in Rethly and Berkes (1963), Isobe *et al*. (2019), and this study (in magnetic latitude). The unlikely candidates in Rethly and Berkes (1963) are in black, other candidates in red. Data from Kirch's report examined in this study are in blue, and the East Asian simultaneous report data in Isobe *et al*. (2019) appears in purple. We have derived the error margin of the actual equatorward boundary of the auroral oval for each case, based on the descriptions of their reported altitude and direction. When neither of its direction nor altitude is described, we have assumed their possible ranges of the elevation angle as 0° − 180° from the poleward horizon.

**5. Possible contributions of the CIR-storms**

The relative proximity between the European sector and the magnetic pole during the MM indicates requirements of weaker magnetic source for the auroral visibility in the European sector and making CIRs from high-speed solar wind more plausible for their sources, while Riley *et al*. (2015) expected ICMEs as their sources based on the auroral visibility in the mid-Europe (Rethly and Berkes, 1963). The CIR-associated storms with their minimal Dst have been empirically known to be as low as − 161 nT (Richardson *et al*., 2006), and the theoretical limit of their minimal Dst has been estimated to

−180 nT (O'Brien and McPherron, 2000). The empirical correlations of the storm intensity by the Dst index, and the equatorward boundary of the auroral ovals in Yokoyama *et al*. (1998), allow us to estimate the equatorward boundary of the auroral oval within the capacity of the CIR-storms as 49.4° ± 1.2° (48.2°−50.6°) MLAT in their footprint of the magnetic field line. The 1.2-times stronger magnetic field at that time does not significantly affect this threshold line, as this would yield the equatorial boundary of the auroral oval for the same-size magnetic storms shift at best ≈ 1° MLAT poleward (Ebihara and Tanaka, 2020).

As the auroral visibility in mid-Europe does not necessarily indicate visibility of the overhead aurora in mid-Europe, it is more likely to expect its visibility in the poleward sky (see *e.g.* Shiokawa *et al*., 2005; Figure 6 of Kimball, 1960; Figure 2 of Hayakawa *et al*., 2018). As such, unless the auroral displays extend beyond the zenith, the actual equatorial boundary of the associated auroral *ovals* stay significantly more poleward than the equatorward boundary of the auroral *visibilities* and yield only weaker storms.

Accommodating these uncertainties, we have re-analysed the candidate aurorae in Rethly and Berkes (1963), which are used in a number of modern studies (*e.g.*, Riley *et al*., 2015; Scafetta and Wilson, 2013; Vaquero and Trigo, 2015). Figure 5 shows the MLATs of reported visibility of candidate aurorae in Rethly and Berkes (1963) and their possible margin of equatorial boundary of the candidate aurorae. Unless the direction is otherwise specified, the variability of their elevation angle was estimated as 0°–180° from the poleward horizon of their observational sites. Here, it is explicitly shown that the reported candidate aurorae stayed within the intensity range of the CIR-storms even if the candidate aurorae in Rethly and Berkes (1963) extended overhead of their observational sites (49.4° ± 1.2° MLAT). Without explicit constraints in their direction, these candidate aurorae were more likely seen in the poleward sky and reduced required intensities of their source storms from the threshold of equatorward extent of the CIR-origin aurorae.

Being derived from the high-speed solar wind, CIR-storms do not necessarily require source sunspots, and can explain the reported absence of simultaneous sunspots in the MM, which has been somewhat puzzling when to expect ICMEs as their sources (Letfus, 2000; Isobe *et al*., 2019). In fact, CIRs have been continuously reported, even during the deep solar minimum in 2008/2009, and its occurrence has not been affected as much as the properties of solar wind and sunspot number (Jian *et al*., 2011). This trend can be extended to the MM with a deeper suppression of the sunspot activity (Riley *et al*., 2015; Usoskin *et al*., 2015; Vaquero *et al*., 2015). In this case, the auroral visibilities in mid-Europe do not necessarily require ICMEs and associated source active regions.

**6. Summary and discussions**

We have analysed the 'fire-signs' reported in Kirch (1681) and documented their observational details. Our analyses show that this event was widely observed in the area around the southern Baltic Sea as well as from middle Germany from 1 LT to local daybreak. The colourations were described as reddish, bluish, golden, and whitish (including silvery). The shapes were described as a combination of glow and rays, whereas the rays were often described as curved with shapes and vertical motions varying with time. The descriptions are consistent with mid-latitude aurorae, rather than with other possible candidates such as fireballs.

Based on the description of the overhead coverage at 55.3° MLAT, the footprint of the magnetic field line for the equatorial boundary of the auroral oval has been estimated as ≈ 56.5° MLAT. With an equatorward boundary of the auroral oval comparable to those of major storms such as the one on 4 August 1972, this storm intensity can be estimated as Dst ≤ −100 nT. Its chronological coincidence with reported sunspot visibility on 20–30 May 1680 tempts us to associate this sunspot with the plausible source of this candidate storm, whereas its intensity indicates CIRs derived from high-speed solar wind as its possible cause. At least, the period of May–June 1680 was peculiar in the MM, hosting both the candidate aurora and the sunspot group.

Its ICME-origin is acceptable, as the sunspot AR is recorded a few days before this candidate aurora, the rate of ICMEs is estimated to be comparable between the recent solar minima (2008/2009 and 1996/1997) and the MM (Owens and Lockwood, 2012), and intensity of this storm was not that extreme (minimal Dst ≤ −100 nT). In fact, despite the empirical preference of ICME occurrence around the maximum, or the declining phase of the solar cycle (Kilpua *et al*., 2015; Lefevre *et al*., 2016), some significant ICMEs have caused extreme geomagnetic storms around the cycle minimum or immediately afterward: e.g., February 1986 storm (minimal Dst = −307 nT; Garcia and Dryer, 1987; WDC for geomagnetism at Kyoto *et al*., 2015), September 1998 storm (minimal Dst index = −207 nT; Daglis *et al*., 2007; WDC for geomagnetism at Kyoto *et al*., 2015), and October 1903 storm (minimum Dst estimate ≈ −531 nT; Hayakawa *et al*., 2020c; see also Ribeiro *et al*., 2016).

This contrasts with the other candidate aurorae in this period, mostly without simultaneous sunspot observations (*e.g.*, Letfus, 2000; Isobe *et al*., 2019). While ICMEs and source sunspots have been expected for the origin of these candidate aurorae (Riley *et al*., 2015), a considerable ratio of the candidate aurorae described during the MM in Rethly and Berkes (1963) were probably misinterpretations, as already clarified by themselves. These discussions probably indicate that the auroral night was even less frequently than previously considered (Vaquero and Trigo, 2015; Riley *et al*., 2015), and reinforces their scenarios of reduced auroral activity during the MM (Eddy, 1976; Siscoe, 1980; Feynman and Gabriel, 1990; Nevanlinna, 1995; Tsurutani *et al*., 2011; Lockwood and Barnard, 2015), despite the relative proximity of the European sector to the magnetic pole than at modern time (Usoskin *et al*., 2015; see also Figure 4). This is consistent with the reconstructed slower solar wind and lower solar wind dynamic pressure during the MM than the recent deep solar cycle minima (Cliver *et al*., 1998; Owens *et al*., 2017). These peculiar conditions more likely made the auroral oval remain at higher latitudes, while also reducing the auroral brightness (Millan *et al*., 2010).

Nevertheless, the auroral activity did not completely cease even in the European sector. Visibility extensions indicate that the intensity remained within a capacity of CIR-storms in addition to previously suggested ICME-storms (Figure 5; *c.f.*, Riley *et al*., 2015). Notably, aurorae were probably less frequent in the European sector than previously considered despite its relative proximity to the magnetic pole (≈ 4° MLAT). Therefore, we conclude that the solar wind was generally slower with quieter auroral activity. However, occasional occurrences of sunspots and coronal holes have sporadically triggered geomagnetic storms and mid-latitude aurorae not only for ICMEs but also in CIRs. Our results likely resolve some puzzling cases of candidate auroral records in mid-Europe without simultaneous sunspot records.

**Acknowledgment**

We thank JSPS Grant-in-Aids JP15H05812, JP17J06954, JP20H00173, and JP20K20918, JSPS Overseas Challenge Program for Young Researchers, the 2020 YLC collaborating research fund, and the research grants for Mission Research on Sustainable Humanosphere from Research Institute for Sustainable Humanosphere (RISH) of Kyoto University and Young Leader Cultivation (YLC) program of Nagoya University. We thank Bayerische Staatsbibliothek and Bibliothek Wolfenbüttel for letting us access the copies of Kirch (1681) and Voigt (1681). HH thanks archivists in l'Observatoire de Paris for letting him access Gottfried Kirch's original manuscript (MS B 3/1-6 Cote Delisle 77), Jean-Pierre Allizart and Valérie Godin for their cordial helps upon his surveys at

Hayakawa et al., 2020, A candidate major solar-terrestrial storm in 1680
*The Astrophysical Journal*, DOI: 10.3847/1538-4357/abb3c2

Hayakawa et al., 2020, A candidate major solar-terrestrial storm in 1680
*The Astrophysical Journal*, DOI: 10.3847/1538-4357/abb3c2